\documentclass[preprint,12pt]{elsarticle}

\usepackage{amssymb}
\usepackage{IEEEtrantools}

\journal{Signal Processing: Image Communication}

\begin{document}

\begin{frontmatter}



\title{Entropy Conserving Binarization Scheme for Video and Image Compression}


\author{Madhur Srivastava}

\address{Department of Biomedical Engineering, Cornell University, Ithaca, New York, USA}

\begin{abstract}
The paper presents a binarization scheme that converts non-binary data into a set of binary strings. At present, there are many binarization algorithms, but they are optimal for only specific probability distributions of the data source. Overcoming the problem, it is shown in this paper that the presented binarization scheme conserves the entropy of the original data having any probability distribution of $m$-ary source. The major advantages of this scheme are that it conserves entropy without the knowledge of the source and the probability distribution of the source symbols. The scheme has linear complexity in terms of the length of the input data. The binarization scheme can be implemented in Context-based Adaptive Binary Arithmetic Coding (CABAC) for video and image compression. It can also be utilized by various universal data compression algorithms that have high complexity in compressing non-binary data, and by binary data compression algorithms to optimally compress non-binary data.

\end{abstract}


\end{frontmatter}


\section{Introduction}
\label{sec:intro}

Data compression is performed in all types of data requiring storage and transmission. It preserves space, energy and bandwidth, while representing the data in most efficient way [1-4]. There are numerous coding algorithms used for compression in various applications [1-3,5-18]. Some of them are optimal [4,19] in all cases, whereas others are optimal for a specific probability distribution of the source symbols. All of these algorithms are mostly applied on $m$-ary data source. However, most of the universal compression algorithms substantially increase their coding complexity and memory requirements when the data changes from binary to $m$-ary source. For instance, in arithmetic coding, the computational complexity difference between the encoder and decoder increases with the number of source symbols [2]. Therefore, it would be beneficial if binarization is perfomed on $m$-ary data source before compression algorithms are applied on it. The process where binarization is followed by compression is most notably found in Context-based Adaptive Binary Arithmetic Coding (CABAC) [20] which is used in H.264/AVC Video Coding Standard [21], High Efficiency Video Coding (HEVC) Standard [22], dynamic 3D mesh compression [23], Audio Video Coding Standard (AVS) [24], Motion Compensated-Embedded Zeroblock Coding (MC-EZC) in scalable video coder [25], multiview video coding [26], motion vector encoding [27-28], and 4D lossless medical image compression [29].

There are many binary conversion techniques which are, or can be used for the binarization process. The most common among all is binary search tree [30-32]. In this, Huffman codeword is used to design an optimal tree [18]. However, there are two limitations to it. First, the probability of all the symbols should be known prior to encoding that may not be possible in all the applications. Although there are methods to overcome the above problem, they come at an additional cost of complexity. For example, binary search tree is updated with the change in incoming symbol probabilities. Second, as with the Huffman coding, the optimality is achieved only when the probability distribution of symbols are in the powers of two. Apart from binary search tree, there are other binarization schemes like unary binarization scheme [20], truncated unary binarization scheme [20], fixed length binarization scheme [20], Golomb binarization scheme [9,20,33-34], among many [5-13,20,30-34]. All of them are optimal for only certain type of symbol probability distributions, and hence, can only conserve the entropy of the data for that probability distribution of the source symbols. Currently, there is no binarization scheme that is optimal for all probability distributions of the source symbols which would result in achieving overall optimal data compression. 

This paper presents a generalized optimal binarization algorithm. The novel binarization scheme conserves the entropy of the data while converting the $m$-ary source data into $m-1$ binary strings. Moreover, the binarization technique is independent of the data type and can be used in any field for storing and compressing data. Furthermore, it can efficiently represent data in the fields which require data to be easily written and read in binary form.

The paper is organized as follows. Section 2 describes the binarization and de-binarization process that will be carried out at the encoder and decoder, respectively. The optimality proof of the binarization scheme is provided in section 3. In section 4, the complexity associated with the binarization process is discussed. Lastly, section 5 concludes by stating the advantages of the presented binarization scheme over others, and its applications.

\section{Binarization and De-binarization}

The binarization of the source symbols is carried out at the encoder using the following two steps: 
\begin{enumerate}
\item A symbol is chosen, and a binary data stream is created by assigning '1' where the chosen symbol is present and '0' otherwise, in the uncompressed data. 
\item The uncompressed data is rearranged by removing the symbol chosen in step 1. 
\end{enumerate}

The two steps are iteratively applied for $m-1$ symbols. It needs to be explicitly emphasized that the binarization of symbol occurs on the previously rearranged uncompressed data and not on the original uncompressed data. Here, the algorithm reduces the uncompressed data size with the removal of binarized symbols from the data, leading to the conservation of entropy. After binarizing every symbol, there are $m-1$ binary data streams corresponding to $m$ source symbols. It is because the $m-1^{th}$ binary string would represent $m-1^{th}$ and $m^{th}$ symbols as '1' and '0', respectively. The binarization scheme demostrated here is optimal i.e., the overall entropy of binarized data streams is equal to the entropy of original data containing $m$-ary source. The proof of optimality is provided in section 3. After binarization, the binarized data streams can be optimally compressed using any universal compression algorithm, including the algorithms that optimally compress only binary data (for example: binary arithmetic coding). 

\begin{table*}[ht!] \footnotesize
\centering
\caption{An example of binarization process}
\begin{tabular}{|c|c|c|c|}
\hline
Binarization Order & \multicolumn{3}{c|}{Iteration} \\ \cline{2-4}
 & First & Second & Third \\
\hline
Data & AABCBACBBACCABACB & BCBCBBCCBCB & CCCCC\\
ABC & 11000100010010100 & 10101100101 & 11111 \\
\hline
Data & AABCBACBBACCABACB & BCBCBBCCBCB & BBBBBB\\
ACB & 11000100010010100 & 01010011010 & 111111 \\
\hline
Data & AABCBACBBACCABACB & AACACACCAAC & CCCCC \\
BAC & 00101001100001001 & 11010100110 & 11111 \\
\hline
Data & AABCBACBBACCABACB & AACACACCAAC & AAAAAA \\
BCA & 00101001100001001 & 00101011001 & 111111 \\
\hline
Data & AABCBACBBACCABACB & AABBABBAABAB & BBBBBB \\
CAB & 00010010001100010 & 110010011010 & 111111 \\
\hline
Data & AABCBACBBACCABACB & AABBABBAABAB & AAAAAA \\
CBA & 00010010001100010 & 001101100101 & 111111 \\
\hline
\end{tabular}
\end{table*}

\begin{table*}[ht!] \footnotesize
\centering
\caption{An example of de-binarization process}
\begin{tabular}{|c|c|c|c|}
\hline
De-binarization Order & \multicolumn{3}{c|}{Iteration} \\ \cline{2-4}
 & First & Second & Third \\
\hline
ABC & 11000100010010100 & AA101A011A00A1A01 & AAB1BA1BBA11ABA1B \\
Data & AA000A000A00A0A00 & AAB0BA0BBA00ABA0B & AABCBACBBACCABACB\\
\hline
ACB & 11000100010010100 & AA010A100A11A0A10 & AA1C1AC11ACCA1AC1 \\
Data & AA000A000A00A0A00 & AA0C0AC00ACCA0AC0 & AABCBACBBACCABACB\\
\hline
BAC & 00101001100001001 & 11B0B10BB1001B10B & AAB1BA1BBA11ABA1B \\
Data & 00B0B00BB0000B00B & AAB0BA0BBA00ABA0B & AABCBACBBACCABACB \\
\hline
BCA & 00101001100001001 & 00B1B01BB0110B01B & 11BCB1CBB1CC1B1CB \\
Data & 00B0B00BB0000B00B & 00BCB0CBB0CC0B0CB & AABCBACBBACCABACB \\
\hline
CAB & 00010010001100010 & 110C01C001CC101C0 & AA1C1AC11ACCA1AC1 \\
Data & 000C00C000CC000C0 & AA0C0AC00ACCA0AC0 & AABCBACBBACCABACB \\
\hline
CBA & 00010010001100010 & 001C10C110CC010C1 & 11BCB1CBB1CC1B1CB \\
Data & 000C00C000CC000C0 & 00BCB0CBB0CC0B0CB & AABCBACBBACCABACB \\
\hline
\end{tabular}
\end{table*}

Table 1 shows the binarization process through an example. A sample input data 'AABCBACBBACCABACB' is considered for the process and as can be seen, it contains three source symbols 'A', 'B', and 'C'. In Table 1, 'Binarization order' states the sequence in which the symbols are binarized. For instance, in 'ABC' binarization order, 'A' is binarized first, followed by 'B', and then finally by 'C'. The row 'Data' shows the uncompressed data available to be binarized after each iteration. Below the 'Data' row is the binarized value of each symbol. As can be seen in each first iteration, the symbol that has be binarized is marked '1', while others are marked '0'. In the next iteration, the symbol that was binarized in the current step is removed from the uncompressed data. Although shown in Table 1, the binarization process does not require to binarize last symbol, because the resultant string contains all '1's that provide no additional information and is redundant. It also needs to be noted that each binarization order results in different sets of binary strings. 

At the decoder, the decoding of the compressed data is followed by de-binarization of $m$-ary source symbol. The order of decoding follows the order of encoding for perfect reconstruction at minimum complexity. With the encoding order information, the de-binarization can be perfectly reconstructed in multiple ways other than the encoding order, but the reordering of sequence after every de-binarization will increase the time as well as the decoder complexity. The de-binarization of the source symbols is also carried out in two steps shown below, and these steps are recursively applied to all the binary data streams representing $m$-ary source symbols: 

\begin{enumerate}
\item Replace '1' with the source symbol in the reconstructed data stream. 
\item Assign the values of the next binary data stream in sequence to the '0's in the reconstructed data stream. 
\end{enumerate}

An example of de-binarization process is shown in Table 2. The de-binarization order follows the same order as of binarization process. In Table 2, the row 'Data' represents the reconstructed data at each iteration. The value '1' is replaced by the symbol to be de-binarized in the respective iteration, while '0's are replaced by the binary string of the next symbol to be de-binarized. Finally, after the last iteration, the original input data 'AABCBACBBACCABACB' is losslessly recovered for all binarization and de-binarization order.

\section{Optimality of Binarization Scheme} 

Let the data source be $Y\in \{Y_1,\, Y_2,\, \dots,\, Y_m\}$, and $X\in \{0,\, 1\}$ be the binary source for each source symbol. The entropy of a $m$-ary source $Y$ is defined as,
\begin{equation} \small
H(Y) = -\sum_{i=1}^{m} p(Y_i) \log p(Y_i) 
\end{equation}
where $p(Y_i)$ is the probability of $Y_i^{th}$ source symbol. Subsequently, the entropy of $m$-ary data source $Y$ with length $N$ is $H(Y^N)$. Similarly, $H(X^N)$ is the entropy of binary source with data length $N$. As explained in the binarization algorithm, the uncompressed data is rearranged after the binarization of the previous symbol/s to ${N(1-\sum_{j=1}^{i-1}p(Y_j))}$ data length i.e., the length of the original data subtracted by the length of all the previously binarized source symbols. Hence, the overall entropy of the $m$ binarized strings is $\sum_{i=1}^{m} H(X_i^{N(1-\sum_{j=1}^{i-1}p(Y_j))})$. Here, $m$ binary strings are considered for mathematical convenience.

To achieve the optimal binarization of $m$-ary source, the entropy of $m$-ary source data must equal the total entropy of binary strings. Therefore,
{\small
\begin{IEEEeqnarray}{rCl} \footnotesize
H(Y^N) & = & \sum_{i=1}^{m} H(X_i^{N(1-\sum_{j=1}^{i-1}p(Y_i))}) \\
N H(Y) & = & \sum_{i=1}^{m} N\left(1-\displaystyle\sum_{j=1}^{i-1} p(Y_i)\right) H(X_i) 
\end{IEEEeqnarray}}

The probability distribution of the binary source $X_i$ is the probability distribution of $m$-ary source $Y_i$ when the first $i-1$ source symbols have already been binarized i.e., removed from the original data. Thus, $H(X_i)$ can be rewritten in terms of $Y_i$ in the following way:

{\small
\begin{IEEEeqnarray}{rCl} \footnotesize
H(Y) & = & \sum_{i=1}^{m} \left(1-\displaystyle\sum_{j=1}^{i-1} p(Y_j)\right) H\left(\frac{p(Y_i)}{1 - \sum_{j=1}^{i-1}p(Y_j)}\right) \\
H(Y) & = & -\sum_{i=1}^{m} \left(1-\displaystyle\sum_{j=1}^{i-1} p(Y_j)\right) \left(\frac{p(Y_i)}{1 - \sum_{j=1}^{i-1}p(Y_j)}\right)\nonumber \\
&& \log \left( \frac{p(Y_i)}{1 - \sum_{j=1}^{i-1}p(Y_j)}\right) - \sum_{i=1}^{m} \left(1-\displaystyle\sum_{j=1}^{i-1} p(Y_j)\right) \nonumber \\
&&  \left[\frac{1 - \sum_{j=1}^{i}p(Y_j)}{1 - \sum_{j=1}^{i-1}p(Y_j)} \log \left( \frac{1 - \sum_{j=1}^{i}p(Y_j)}{1 - \sum_{j=1}^{i-1}p(Y_j)}\right)\right]
\end{IEEEeqnarray}
\begin{IEEEeqnarray}{rCl} \footnotesize
H(Y) & = & -\sum_{i=1}^{m} p(Y_i) \log \left( \frac{p(Y_i)}{1 - \sum_{j=1}^{i-1}p(Y_j)}\right) \nonumber \\
&& -\> \sum_{i=1}^{m} (1 - \sum_{j=1}^{i}p(Y_j)) \log \left( \frac{1 - \sum_{j=1}^{i}p(Y_j)}{1 - \sum_{j=1}^{i-1}p(Y_j)}\right)
\end{IEEEeqnarray}
\begin{IEEEeqnarray}{rCl} \footnotesize
H(Y) & = & -\sum_{i=1}^{m} p(Y_i) \log \left( \frac{p(Y_i)}{\sum_{j=i}^{m}p(Y_j)}\right) \nonumber \\   
&& -\> \sum_{i=1}^{m} (\sum_{j=i+1}^{m}p(Y_j)) \log \left( \frac{\sum_{j=i+1}^{m}p(Y_j)}{\sum_{j=i}^{m}p(Y_j)}\right)
\end{IEEEeqnarray}
\begin{IEEEeqnarray}{rCl} \footnotesize
H(Y) & = & -\sum_{i=1}^{m} \left(p(Y_i) \log p(Y_i) - p(Y_i) \log \left(\sum_{j=i}^{m}p(Y_j)\right)\right) \nonumber \\
&& -\> \sum_{i=1}^{m}  \left(\sum_{j=i+1}^{m}p(Y_j)\right) \log \left(\sum_{j=i+1}^{m}p(Y_j)\right) \nonumber \\
&& +\> \sum_{i=1}^{m} \left(\sum_{j=i+1}^{m}p(Y_j)\right) \log \left(\sum_{j=i}^{m}p(Y_j)\right)
\end{IEEEeqnarray}
\begin{IEEEeqnarray}{rCl} \footnotesize
H(Y) & = & -\sum_{i=1}^{m} p(Y_i) \log p(Y_i) \nonumber \\
&& -\> \sum_{i=1}^{m} \left(\sum_{j=i+1}^{m}p(Y_j)\right) \log \left(\sum_{j=i+1}^{m}p(Y_j)\right) \nonumber \\
&& +\> \sum_{i=1}^{m} \left(\sum_{j=i}^{m}p(Y_j)\right) \log \left(\sum_{j=i}^{m}p(Y_j)\right)
\end{IEEEeqnarray}
\begin{IEEEeqnarray}{rCl} \footnotesize
H(Y) & = & -\sum_{i=1}^{m} p(Y_i) \log p(Y_i) \nonumber \\
&& +\> \left(\sum_{j=1}^{m}p(Y_j)\right) \log \left(\sum_{j=1}^{m}p(Y_j)\right)\\
H(Y) & = & -\sum_{i=1}^{m} p(Y_i) \log p(Y_i) + 1 \log 1 \\
H(Y) & = & -\sum_{i=1}^{m} p(Y_i) \log p(Y_i)
\end{IEEEeqnarray}}
The reduction of equation 2 to equation 12 (also equation 1) proves that the binarization scheme preserves entropy for any $m$-ary data source. 

\section{Computational Complexity of Binarization Scheme}

The computational complexity of the presented method is the linear function of the input data length. The binarization and de-binarization process only acts as a filter, assigning or replacing 0's and 1's, respectively, for an occurrence of a source symbol without any additional table or calculation, that is created or performed for the other binarization techniques. Suppose, the length of input data is $N$, $m$ is the number of source symbols, and $Y$ is the source. For the first symbol, the length of the binary string would be $N$. The length of binary string for the second symbol would be the length of all the symbols, except the first symbol (see Table 1). Likewise, the length of $i^{th}$ binary string would be the length all symbols yet to be binarized. Mathematically, the length can be written as {\small $N\left(1-\displaystyle\sum_{j=1}^{i-1} p(Y_i)\right)$}, where $p(Y_i)$ is the probability of $i^{th}$ symbol. The total number of binary assignment would be {\small $\sum_{i=1}^{m}N\left(1-\displaystyle\sum_{j=1}^{i-1} p(Y_i)\right)$}. As can be seen, the computational complexity of the binarization and de-binarization process is linear in terms of the input data length.

\section{Conclusion: Advantages and Applications}

The proposed binarization scheme has the following advantages over others. Firstly, it is optimal for every data set. As proved and shown in this paper, the binarization scheme conserves entropy of $m$-ary data source. Secondly, the proposed method eliminates the need for knowing the source symbols at all. It works optimally without the knowledge of source because the binarization of the source symbols can occur in any order as shown Table 1, and all orders conserve $m$-ary source entropy, which can be inferred from the derivation shown in section 3. Thirdly, adding to the previous point, the coding is independent of the occurrence of the source symbols. In other words, any source symbol can be encoded in any order subject to the constraint that decoding is performed in the same order. The optimality is independent of the source order in the data set. Fourthly, unlike variable length codes, there is no need to know the probability distribution of the source symbols beforehand. It can be updated as the symbols occur. However, even without the knowledge of probability distribution, the presented method is optimal. Lastly, it has low complexity that is feasible for practical data compression. 

One of the immediate usage of the presented binarization technique is in CABAC used in video and image compression. In addition, CABAC with the proposed binarization scheme can potentially replace Context-based Adaptive Arithmetic Coding used in various image compression standards [35], including JPEG2000 [36]. Furthermore, the binarization scheme can be applied to all the universal compression algorithms that have less complexity and resource requirements for binary data, than $m$-ary data.




\begin{thebibliography}{10}
\expandafter\ifx\csname url\endcsname\relax
  \def\url#1{\texttt{#1}}\fi
\expandafter\ifx\csname urlprefix\endcsname\relax\def\urlprefix{URL }\fi
\expandafter\ifx\csname href\endcsname\relax
  \def\href#1#2{#2} \def\path#1{#1}\fi

\bibitem{1}
T.~Bell, J.~Cleary, I.~Witten, in: Text Compression, Prentice-Hall, Englewood
  Cliffs, New Jersey, 1990.

\bibitem{2}
K.~Sayood, in: Lossless Compression Handbook, Academic Press, Boston, 2003.

\bibitem{3}
D.~Salomon, in: Data Compression, Springer Verlag, New York, 2000.

\bibitem{4}
T.~Cover, J.~Thomas, in: Elements of Information Theory, Wiley-Interscience,
  New Jersey, 2006.

\bibitem{5}
V.~Levenstein, On the redundancy and delay of separable codes for the natural
  numbers, Problems of Cybern. 20 (1968) 173--179.

\bibitem{6}
P.~Elias, Universal codeword sets and representations of the integers, IEEE
  Trans. Inform. Theory 21 (1975) 194--203.

\bibitem{7}
S.~Even, M.~Rodeh, Economical coding of commas between strings, Commun. of the
  ACM 21 (1978) 315--317.

\bibitem{8}
R.~Rice, Some practical universal noiseless coding techniques, Jet Propulsion
  Laboratory Publication 79.

\bibitem{9}
S.~Golomb, Run-length encodings, IEEE Trans. Inform. Theory 12 (1966) 399--401.

\bibitem{10}
A.~Apostolico, A.~Fraenkel, Robust transmission of unbound strings using
  fibonacci representations, IEEE Trans. Inform. Theory 33 (1987) 238--245.

\bibitem{11}
A.~Fraenkel, S.~Klein, Robust universal complete codes for transmission and
  compression, Discrete Appl. Math. 64 (1996) 31--55.

\bibitem{12}
P.~Fenwick, Ziv-lempel encoding with multi-bit flags, Proc. Data Compression
  Conf. (1993) 138--147.

\bibitem{13}
D.~Huffman, A method for the construction of minimum redundancy codes, Proc. of
  the I.R.E. 40 (1952) 1098--1101.

\bibitem{14}
J.~Ziv, A.~Lempel, A universal algorithm for sequential data compression, IEEE
  Trans. Inform. Theory 23 (1977) 337--343.

\bibitem{15}
J.~Ziv, A.~Lempel, Compression of individual sequences via variable-rate
  coding, IEEE Trans. Inform. Theory 24 (1978) 530--536.

\bibitem{16}
T.~Welch, A technique for high-performance data compression, IEEE Computer 17
  (1984) 8--19.

\bibitem{17}
R.~Gallager, Variations on a theme by huffman, IEEE Trans. Inform. Theory 24
  (1978) 668--674.

\bibitem{18}
G.~Langdon, An introduction to arithmetic coding, IBM J. of Research and
  Develop. 28 (1984) 135--149.

\bibitem{19}
C.~Shannon, A mathematical theory of communication, Bell Syst. Tech. J. 27
  (1948) 379--423.

\bibitem{20}
D.~Marpe, H.~Schwarz, T.~Wiegand, Context-based adaptive binary arithmetic
  coding in the h.264/avc video compression standard, IEEE Trans. Circuits and
  Syst. Video Technol. 13 (2003) 620--636.

\bibitem{21}
T.~Wiegand, G.~Sullivan, G.~Bjontegaard, A.~Luthra, Overview of the h.264/avc
  video coding standard, IEEE Trans. Circuits and Syst. Video Technol. 13
  (2003) 560--576.

\bibitem{22}
W.-J.~H. G.J.~Sullivan, J.~Ohm, T.~Wiegand, Overview of the high efficiency
  video coding (hevc) standard, IEEE Trans. Circuits and Syst. Video Technol.
  22 (2012) 1649--1668.

\bibitem{23}
K.~Muller, A.~Smolic, M.~Kautzner, P.~Eisert, T.~Wiegand, Predictive
  compression of dynamic 3d meshes, in: Proc. IEEE Int. Conf. on Image
  Process., Genoa, Italy, Vol.~1, 2005, pp. 621--624.

\bibitem{24}
L.~Zhang, X.~Wu, N.~Zhang, W.~Gao, Q.~Wang, D.~Zhao, Context-based arithmetic
  coding reexamined for dct video compression, in: Proc. IEEE Int. Symp. on
  Circuits and Syst., 2007, pp. 3147--3150.

\bibitem{25}
Y.~Wu, J.~Woods, Scalable motion vector coding based on cabac for mc-ezbc, IEEE
  Trans. Circuits and Syst. Video Technol. 17 (2007) 790--795.

\bibitem{26}
Y.~Sehoon, A.~Vetro, Rd-optimized view synthesis prediction for multiview video
  coding, in: Proc. IEEE Int. Conf. on Image Process., San Antonio, Texas, USA,
  Vol.~1, 2007, pp. 209--212.

\bibitem{27}
R.~Kordasiewicz, M.~Gallant, S.~Shirani, Encoding of affine motion vectors,
  IEEE Trans. on Multimedia 9 (2007) 1346--1356.

\bibitem{28}
A.~Golwelkar, J.~Woods, Motion-compensated temporal filtering and motion vector
  coding using biorthogonal filters, IEEE Trans. Circuits and Syst. Video
  Technol. 17 (2007) 417--428.

\bibitem{29}
V.~Sanchez, P.~Nasiopoulos, R.~Abugharbieh, Efficient 4d motion compensated
  lossless compression of dynamic volumetric medical image data, in: Proc. IEEE
  Int. Conf. on Acoustics, Speech and Signal Process., Las Vegas, Nevada, USA,
  2008, pp. 549--552.

\bibitem{30}
P.~Gill, M.~Wright, W.~Murray, in: Practical Optimization, Academic Press, New
  York, 1982.

\bibitem{31}
J.~Rice, in: Numerical Methods, Software and Analysis, McGraw-Hill, New York,
  1983.

\bibitem{32}
W.~Press, S.~Teukolsky, W.~Vetterling, B.~Flannery, in: Numerical Recipes in C:
  The art of Scientific Computing, Cambridge Univ. Press, Cambridge, UK, 1993.

\bibitem{33}
J.~Teuhola, A compression method for clustered bitvectors, Inform. Process.
  Lett. 7 (1978) 308--311.

\bibitem{34}
R.~Gallager, D.~Voorhis, Optimal source codes for geometrically distributed
  integer alphabets, IEEE Trans. Inform. Theory 21 (1975) 228--230.

\bibitem{35}
K.~Ong, W.~Chang, Y.~Tseng, Y.~Lee, C.~Lee, A high throughput low cost
  context-based adaptive arithmetic codec for multiple standards, in: Proc.
  IEEE Int. Symp. on Circuits and Syst., Vol.~1, 2002, pp. 872--875.

\bibitem{36}
K.~Ong, W.~Chang, Y.~Tseng, Y.~Lee, C.~Lee, A high throughput context-based
  adaptive arithmetic codec for jpeg2000, in: Proc. IEEE Int. Symp. on Circuits
  and Syst., Vol.~4, 2002, pp. 133--136.

\end{thebibliography}





\end{document}